\documentclass[preprint,showpacs,preprintnumbers,amsmath,amssymb,superscriptaddress]{revtex4}

\usepackage{graphicx}
\usepackage{dcolumn}
\usepackage{bm}

\begin{document}

\preprint{APS/123-QED}

\title{Charge manipulation and imaging of the Mn acceptor state in GaAs by Cross-sectional Scanning Tunneling Microscopy}

\author{A.M. Yakunin}
 \affiliation{COBRA Inter-University, Eindhoven University of Technology, P.O.Box 513, NL-
5600MB Eindhoven, The Netherlands}
\altaffiliation{Corresponding author: Andrei M. Yakunin\\E-mail: a.m.yakunin@tue.nl}

 \author{A.Yu. Silov}
 \affiliation{COBRA Inter-University, Eindhoven University of Technology, P.O.Box 513, NL-
5600MB Eindhoven, The Netherlands}
 \author{P.M. Koenraad}
 \affiliation{COBRA Inter-University, Eindhoven University of Technology, P.O.Box 513, NL-
5600MB Eindhoven, The Netherlands}
\author{J.H. Wolter}
\affiliation{COBRA Inter-University, Eindhoven University of Technology, P.O.Box 513, NL-
5600MB Eindhoven, The Netherlands}
\author{W. Van Roy}
\affiliation{IMEC, Kapeldreef 75, B-3001 Leuven, Belgium}
\author{J. De Boeck}
\affiliation{IMEC, Kapeldreef 75, B-3001 Leuven, Belgium}

\date{11 December 2003}

\begin{abstract}
An individual Mn acceptor in GaAs is mapped by Cross-sectional Scanning Tunneling Microscopy (X-STM) at room temperature and a strongly anisotropic shape of the acceptor state is observed. An acceptor state manifests itself as a cross-like feature which we attribute to a valence hole weakly bound to the Mn ion forming the (Mn$^{2+}3d^5+hole$) complex. We propose that the observed anisotropy of the Mn acceptor wave-function is due to the d-wave present in the acceptor ground state.
\end{abstract}

\maketitle

\subsection{Introduction}
By now it is generally accepted that ferromagnetism in diluted magnetic semiconductors such as Ga$_{1-x}$Mn$_{x}$As is driven by the valence band states. A detailed investigation of the hole distribution around magnetic dopants is therefore of essential importance \cite{Ohno}.

Deep acceptors in III/V semiconductors such as Mn$_{Ga}$ have been studied intensively for at least 30 years with a great variety of different techniques such as piezo-spectroscopy, hot photoluminescence, electron magnetic resonance, etc. However, hardly any information has been obtained on the electronic configuration at the atomic scale. Scanning tunneling microscopy is an ideal technique for the spatial investigation of complicated electronic structures, such as long-range interaction between an impurity and the host crystal. STM study of shallow acceptors such as Zn and Cd reveals that the behavior of doping atoms is still not fully understood \cite{Zn,ZnCd}. Recently, isolated manganese acceptors in their ionized state in GaAs and Mn in low temperature grown high concentration Ga$_{0.995}$Mn$_{0.005}$As films have been studied by X-STM \cite{MnIonized,HighConc}. In this work we present a detailed investigation of an isolated manganese acceptor at an atomic scale where we have used the STM tip as a tool to manipulate the manganese acceptor charge state $A^{-}/A^{0}$.We have imaged the Mn dopant by X-STM in both these charge states.

\subsection{Samples}
The measurements were performed on MBE grown samples. GaAs layers of 1200 nm thickness doped with Mn at a concentration of about $3\times10^{18}~cm^{-3}$ were grown on an intrinsic (001) GaAs substrate. The growth temperature was chosen as 580~$^o$C in order to prevent the appearance of the structural defects such as As antisites which would shift the position of the sample Fermi level. The concentration of the Mn dopants should be low enough to prevent them from interacting with each other and forming an impurity band. The samples we used were non-conducting below $77~K$. The X-STM experiments were performed at room temperature on an \emph{in-situ} cleavage induced mono-atomically flat (110) surface in UHV ($P<2\times10^{-11}~torr$).

\subsection{Experiment}
\begin{figure}
  \includegraphics{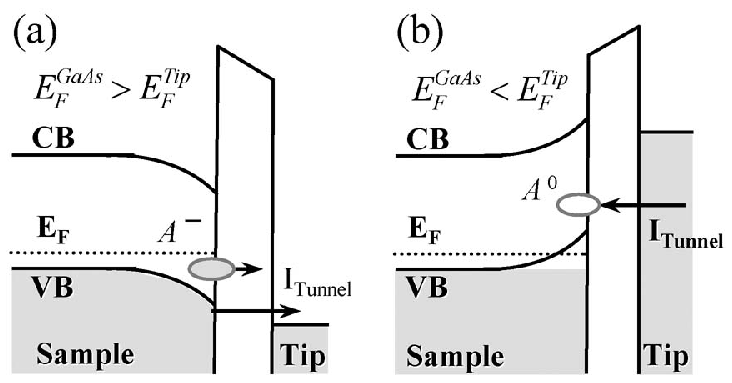}
  \caption{Energy band diagram illustrating the tunneling process between tungsten tip and GaAs~(110) surface in the presence of a tip induced band bending: a)~negative sample bias, filled states tunneling; b)~positive sample bias, empty states tunneling.}\label{Fig1ab}
\end{figure}

The key point of our experiment is that the population of the acceptor state is influenced by the tip-induced band bending on the surface of the semiconductor. The band bending along with the population of the Mn acceptor state can be manipulated by the voltage applied between the STM tip and the sample (figure \ref{Fig1ab}). We have observed and studied the voltage dependent appearance of the Mn in the STM image.

\begin{figure}
  \includegraphics{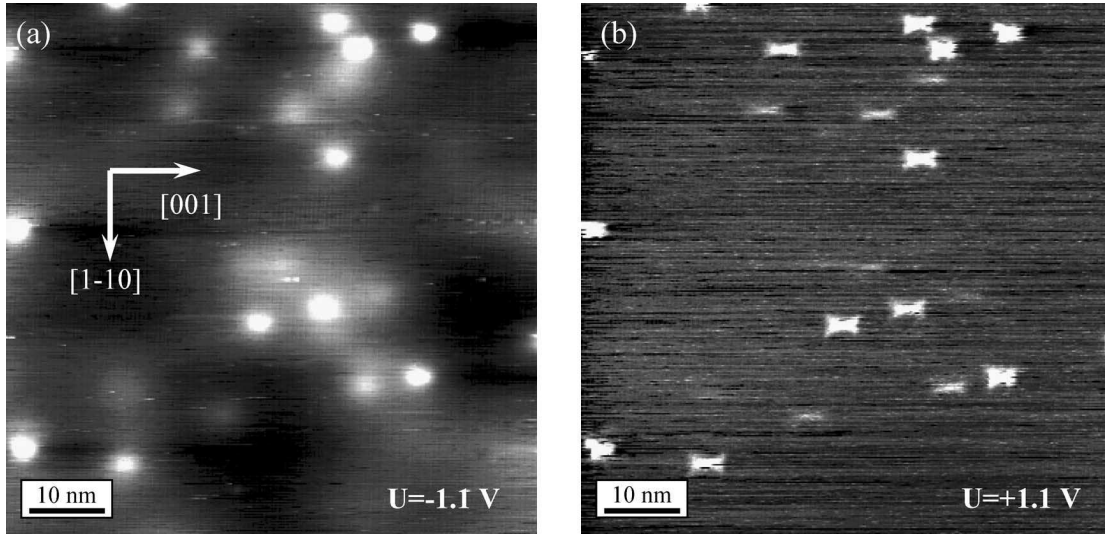}
  \caption{Constant-current STM images $70\times70~nm^2$ of the same area. The Mn dopants are either in their (a) ionized (sample voltage is~$-1.1~V$) or (b) neutral (sample voltage is~$+1.1~V$) charge state. Both images display electronic contrast. In the image (a) the contrast is dominated by Coulomb field influence of the negatively charged Mn$_{Ga}$ ions on the neighboring states of the valence band available for the tunneling. In the image~(b) the bright anisotropic feature appears as soon as the acceptor state is available for the tunneling. The brightest ones have strong electronic contrast, as big as 9~\AA. The details of the feature are still visible without atomic resolution.}\label{Fig2ab}
\end{figure}

\begin{figure}
  \includegraphics{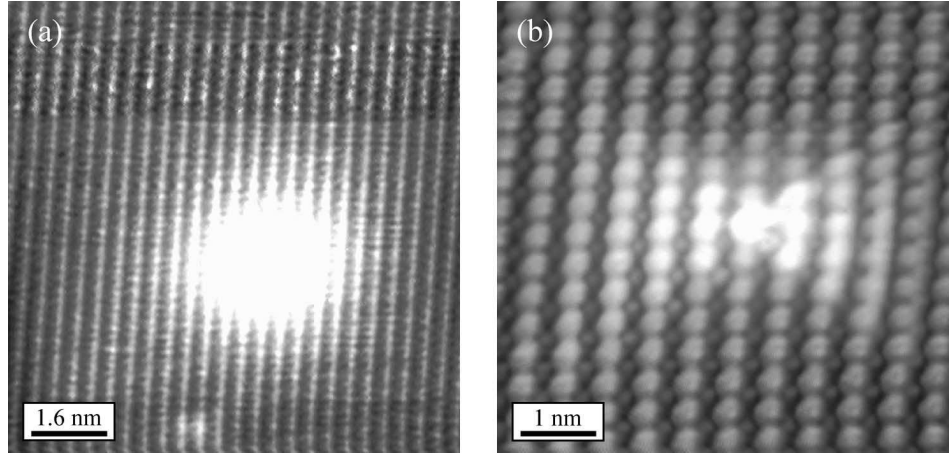}
  \caption{a) $10\times10~nm^2$ constant-current image taken at~$-0.7~V$ shows round isotropic elevation induced by ionized manganese. b)~$6\times6~nm^2$ constant-current image taken at~$+0.6~V$. Due to the particular tunneling conditions a distinctive resolution is achieved and both As (big round features) and Ga (small round features) are visible. The symmetry and the contrast of the feature indicate that Mn dopant is situated in the 3$^{rd}$ subsurface atomic layer.}\label{Fig3ab}
\end{figure}

\begin{figure}
  \includegraphics{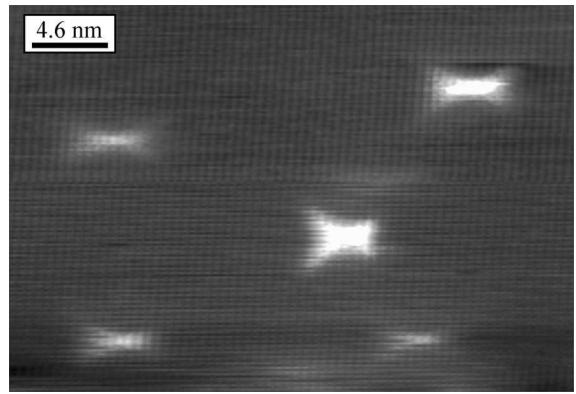}
  \caption{Image displays 5 several Mn dopants in the neutral configuration situated in different subsurface atomic layers. The weaker the contrast of the feature the deeper it is situated under the surface. The image is taken at~$+0.8~V$.}\label{Fig4}
\end{figure}

\begin{figure}
  \includegraphics{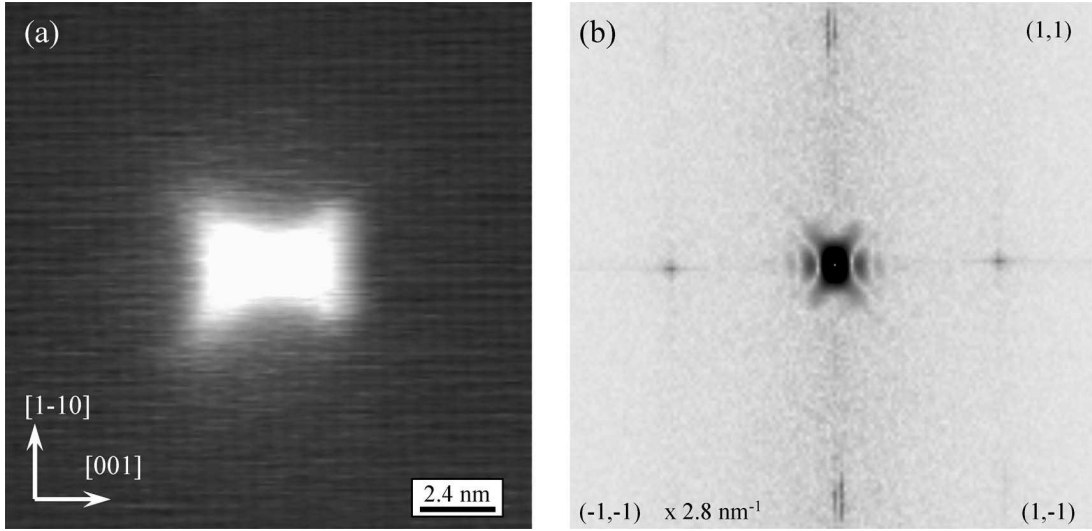}
  \caption{a)~$18\times18~nm^2$ constant-current image taken at $+0.9~V$ shows spatial anisotropy of the feature induced by manganese in the neutral configuration. The electronic contrast of the feature is about 9~\AA. b)~Fourier transform of the image (a).}\label{Fig5ab}
\end{figure}

\begin{figure}
  \includegraphics{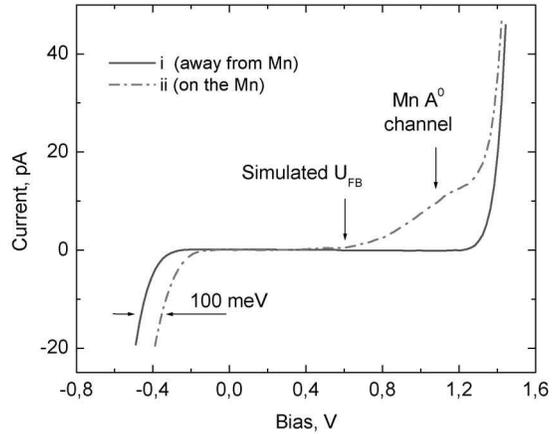}
  \caption{Local Tunneling Spectroscopy \cite{Fenstra}. The set point has been selected at~$+1.6~V$ and $100~pA$. At each point of the image an $I(V)$ spectrum is taken. At the set point bias the electronic contrast of the cross-like feature is almost absent, thus the tip-sample distance remains constant for each spectra. Spectrum~(i), taken at a clean GaAs~(110) surface away from Mn, displays a band gap of about $1.5~eV$; (ii)~taken on the Mn. In the area of the cross like feature an extra current-channel appears in the band gap at low positive voltages.}\label{Fig6}
\end{figure}

At negative sample-bias Mn is found in its ionized configuration. At high negative bias ($U<-0.5~V$) it appears as an isotropic round elevation which is a consequence of the influence of the $A^-$ ion Coulomb field on the valence band states (figures \ref{Fig2ab}a and \ref{Fig2ab}b). This is in agreement with a recent study of the isolated Mn in GaAs in the ionized configuration~\cite{MnIonized}. We have found that at a positive bias Mn is neutral. At high positive voltages cross-like structure disappears above $1.5~V$, when the conduction band empty states dominate the tunnel current (see positive branch of the $I(V)$ characteristic in the figure \ref{Fig6}). This is the direct evidence that no extra charge is confined by the Mn under these tunneling conditions and the dopant is neutral. At low positive voltage where the tip Fermi level is below the conduction band edge Mn shows up as a highly anisotropic cross-like feature (figures \ref{Fig2ab}b, \ref{Fig3ab}b, \ref{Fig4} and \ref{Fig5ab}a). The same feature can bee seen in a narrow range of low voltages ($-0.5<U<-0.4~V$), when the valence band bulk states are not yet involved in the tunneling. The values of the range depend on the tip-sample distance.

The cross-like feature manifests itself in the local tunneling $I(V)$ spectroscopy at low voltages. It appears as an empty- or filled state current channel in the band gap of GaAs depending on applied positive or negative bias, respectively. Thus the mapping of the Mn acceptor state in the filled (empty) states mode is realized by electron (hole) injection into the $A^{0}~(A^{-})$ state. In the tunneling $I(V)$ spectroscopy manganese induced $A^{0}$ channel appears above the flat band potential $U_{FB}$ and is available for tunneling in a wide range of voltages above $U_{FB}$. Our estimated value of $U_{FB}$ is about $+0.6~eV$. Imaging under these conditions provides high contrast of the cross-like features. The contrast can be as high as 9~\AA~(figure~\ref{Fig5ab}a). The observed ionization energy which is determined from the shift of the $I(V)$ spectrum at the negative bias corresponds to an Mn acceptor binding energy $E_{a}=0.1~eV$.

The concentration of the dopants we observe in the STM corresponds to intentional $3\times10^{18}~cm^{-3}$ doping level. All of the observed dopants can be found either in the ionized $A^-$ or the neutral $A^0$ charge state depending on either negative or positive sample bias respectively. In the experiment we identify manganese atoms situated in 6 different layers under the surface. In order to determine the actual position of the Mn dopants we have analyzed the intensity of the electronic contrast of the Mn related features. Based on the symmetry of the cross-like feature superimposed on the surface lattice it is possible to distinguish whether the dopant is located in an even or odd subsurface layer. We found that at any depth the shape of the cross-like feature has clearly twofold symmetry and is elongated along the [001] direction. The cross-like features induced by deeper laying dopants are more elongated in [001] direction as shown in figure 4.
The cross-like feature is weakly asymmetric with respect to a [1-10] surface direction. The symmetry of the (110) surface as well as buckling of the surface atoms can be the cause of this distortion (figures. \ref{Fig3ab}b, \ref{Fig4} and \ref{Fig5ab}a). In the figure \ref{Fig5ab}a the left-hand part extends further away than the right-hand one. The orientation of the larger part is the same as that of the triangular features induced by Zn and Cd dopants in GaAs. In the area of the smaller part there is a considerable atomic corrugation change (see figure \ref{Fig3ab}b). The observed apparent shift of atomic rows in the [001] direction can be as much as 2.5~\AA~even when the dopant is situated as deep as in third sub-surface layer. Since this corrugation change appears only at low positive voltages and is not observed at higher voltages we conclude that it has an electronic origin and is not related to a reconstruction or considerable lateral displacement of atoms.
The anisotropy of the acceptor state is most evident in a reciprocal-space representation. In the figure 5b we present the Fourier spectra of the experimental image shown in figure \ref{Fig5ab}a. Note the presence of the satellite harmonics which arise from the steep fall off the wave-function in the [001] direction.

\subsection{Discussion}
Depending on the concentration and a host crystal, substitutional Mn$_{III}$ can be found in three different electronic configurations. Firstly, it can be an ionized acceptor $A^-$ in the Mn$^{2+}$3$d^5$ electronic state. Secondly, in the neutral acceptor state A 0 formed by a negatively charged core weakly binding a valence hole forming a (Mn$^{2+}$3$d^5+hole$) complex. Thirdly, a neutral configuration 3$d^4$ can occur if a hole enters into the $d$-shell of Mn \cite{Sapega}.
According to \cite{Sapega} at low concentration the neutral configuration of the manganese acceptor is (Mn$^{2+}$3$d^5+hole$). We suggest that the observed anisotropy is due to the presence of the d-wave in the acceptor ground state. We believe that the observed anisotropy of the Mn acceptor on the surface is a bulk phenomenon and that the symmetry of the cross-like feature is dictated by the $\Gamma_8$ point symmetry and the cubic properties of the III/V crystal as described in \cite{PRL_Mn,Pikus,Schechter,Kohn,Baldereschi}.

\subsection{Conclusions}
We studied the electronic structure of a single Mn acceptor in GaAs in both its ionized and neutral configurations at atomic scale. The neutral configuration of the observed Mn acceptor in GaAs was identified as a complex of a negative core and a weakly bound valence hole (Mn$^{2+}$3$d^5+hole$). We have mapped the charge distribution of the bound hole in the vicinity of a cleavage-induced surface. Our experiments reveal an anisotropic spatial structure of the hole, which appears in STM images as a cross-like feature.

\subsection{Acknowledgements}
This work was supported by the Dutch Foundation for Fundamental Research on Matter (FOM), Belgian Fund for Scientific Research Flanders (FWO) and EC GROWTH project FENIKS (G5RD-CT-2001-00535).
\newpage


\newpage

\end{document}